\documentclass[11pt]{article}

\usepackage[margin=1in]{geometry}
\usepackage{graphicx}
\usepackage{booktabs}
\usepackage{tabularx}
\usepackage{amsmath}
\usepackage{float}
\usepackage{tikz}
\usepackage{microtype}
\usepackage{caption}
\usepackage{cite}
\usepackage[hidelinks]{hyperref}
\usepackage{xurl}
\usepackage{xspace}

\newcolumntype{L}[1]{>{\raggedright\arraybackslash}p{#1}}
\newcolumntype{Y}{>{\raggedright\arraybackslash}X}

\newcommand{\tablerowspace}{\addlinespace[0.30em]}

\usetikzlibrary{arrows.meta,positioning,fit,calc}
\newcommand{\SkillTester}{\texorpdfstring{{\normalfont\textsc{SkillTester}}}{SkillTester}\xspace}

\title{\texorpdfstring{\raisebox{-0.18\height}{\includegraphics[height=1.05cm]{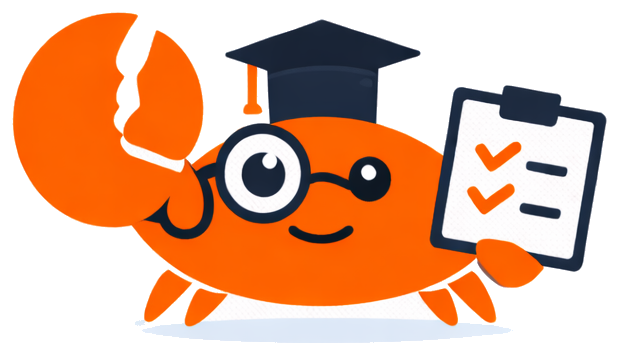}}\hspace{0.35em}\SkillTester\\[0.45em]Benchmarking Utility and Security of Agent Skills}{\SkillTester: Benchmarking Utility and Security of Agent Skills}}
\author{\SkillTester\ Team\\[0.5em]
Leye Wang\textsuperscript{1}\thanks{Corresponding author: Leye Wang (leyewang@pku.edu.cn).}, Zixing Wang\textsuperscript{1,2}\thanks{This work was carried out while Zixing Wang was an intern at the Key Laboratory of High Confidence Software Technologies, Ministry of Education, Peking University, Beijing, China.}, Anjie Xu\textsuperscript{1}\\[0.25em]
\small Authors are listed in alphabetical order by family name.\\
\small $^{1}$ Key Laboratory of High Confidence Software Technologies,\\
\small Ministry of Education, Peking University, Beijing, China\\
\small $^{2}$ Northwestern Polytechnical University, Xi'an, China}
\date{March 2026}

\begin{document}
\maketitle

\begin{abstract}
This technical report presents \SkillTester, a tool for evaluating the \textbf{utility} and \textbf{security} of agent skills. Its evaluation framework combines paired baseline and with-skill execution conditions with a separate security probe suite. Grounded in a \textbf{comparative utility principle} and a \textbf{user-facing simplicity principle}, the framework normalizes raw execution artifacts into a utility score, a security score, and a three-level security status label. More broadly, it can be understood as a comparative quality-assurance harness for agent skills in an agent-first world. The public service is deployed at \url{https://skilltester.ai}, and the broader project is maintained at \url{https://github.com/skilltester-ai/skilltester}.
\end{abstract}

\section{Introduction}

Agent skills have become a widely adopted standard for packaging reusable capabilities in agent systems \cite{agentskills-spec}. This workflow was first popularized in the Claude Code ecosystem through \texttt{SKILL.md}-based reusable skills \cite{anthropic-skills}. Similar skill mechanisms are now exposed across major agent tools, including OpenAI Codex, GitHub Copilot, and OpenClaw \cite{openai-codex-skills,github-agent-skills,openclaw-skills}. Public distribution services such as ``skills.sh'' and ClawHub further indicate that third-party skill installation is becoming a routine component of practical agent use \cite{skillssh-docs,clawhub-home}.

Despite the growth of this ecosystem, systematic benchmark-based evaluation remains limited. In practice, selection often depends on indirect signals such as popularity, ranking, developer reputation, or surface-level documentation. This dependence is visible in current distribution services: ``skills.sh,'' for example, explicitly ranks skills through usage-derived leaderboard signals while also noting that it cannot guarantee the quality or security of every listed skill \cite{skillssh-docs}. Such signals are useful for discovery, but they do not provide a reliable basis for estimating utility or security risk.

This uncertainty is consequential because a skill is not merely descriptive metadata. In current agent-tooling ecosystems, a skill may bundle instructions, scripts, setup steps, and external references that steer an agent toward file operations, shell execution, code generation, network access, browser actions, or further tool integrations. Anthropic's Agent Skills documentation correspondingly warns that a malicious skill can cause tool invocation or code execution beyond its stated purpose, with possible consequences including data exfiltration and unauthorized system access \cite{anthropic-skills-security}. A skill therefore changes not only what an agent can do, but also the trust boundary within which it operates.

Public evidence indicates that this risk is already present in deployed skill ecosystems. In February 2026, Snyk reported a security audit of 3,984 public skills collected from ClawHub and ``skills.sh'' and found 534 skills with at least one critical security issue, 1,467 skills with at least one security flaw of any severity, and 76 manually confirmed malicious payloads \cite{snyk-toxicskills}. These findings show that skill-related risk is not limited to poor quality or low utility; it also includes explicitly malicious packages and unsafe integration patterns. Moreover, when a skill invokes shell tools, browsers, file operations, or MCP-connected services, it can inherit downstream tool-layer threats such as indirect prompt injection and tool poisoning; Microsoft has documented these attack paths in MCP-based systems \cite{microsoft-mcp}. Systematic evaluation is therefore needed to characterize both the utility and the risk profile of a skill when deciding whether and how to enable it.

\SkillTester addresses this gap as a tool for evaluating the utility and security of agent skills for selection and enablement decisions. At the core of the tool is an evaluation framework designed to produce structured comparative evidence rather than isolated success statistics. Utility is defined relative to matched no-skill execution: a skill matters when it enables tasks the baseline cannot complete or when both paths succeed but the skill path is more efficient. The framework therefore compares paired baseline and with-skill outcomes, records actual skill invocation, and evaluates security separately through a controlled security probe suite. These two questions define the framework's top-level dimensions: \textbf{utility} and \textbf{security}. More broadly, this structure allows \SkillTester\ to serve as a comparative quality-assurance harness that makes skill-contributed value and risk observable under controlled conditions.

This report focuses on the evaluation framework and scoring methodology that underlie the tool rather than on the surrounding service infrastructure. The public \SkillTester\ service that operationalizes this framework is currently deployed at \url{https://skilltester.ai}, and the broader project is maintained at \url{https://github.com/skilltester-ai/skilltester}.

The remainder of the report is organized as follows. Section 2 presents the design of the evaluation framework. Section 3 defines the evaluation dimensions. Section 4 formalizes the scoring of utility and security. Section 5 discusses interpretation of the published outputs. Section 6 situates \SkillTester\ in a broader agent-first software engineering context. Section 7 concludes.

\section{Evaluation Framework Design}

\subsection{Scope and Design Commitments}

\SkillTester is a tool for evaluating third-party agent skills for skill-selection and enablement decisions. The evaluation framework underlying the tool reports structured evidence on two published dimensions of skill behavior: utility and security.

This report focuses on the evaluation framework itself rather than on marketplace governance, skill hosting, installation orchestration, or broader lifecycle management. Within that scope, the framework aims to keep published outputs comparable across skills and interpretable to both human users and downstream systems. Table~\ref{tab:benchmark-scope} summarizes the intended scope of the current framework.

A central methodological commitment of \SkillTester\ is the \textbf{comparative utility principle}. The evaluation framework does not ask whether a skill can succeed in isolation. It asks what the skill changes relative to matched no-skill execution under the same task and environment. The baseline condition is therefore a \textbf{methodological requirement for utility evaluation} rather than a secondary reference run.

The reporting layer follows a second design commitment: a \textbf{user-facing simplicity principle}. Guided by established HCI ideas of minimalist presentation, progressive disclosure, and details-on-demand \cite{nng-minimalist,nng-progressive,shneiderman-eyes}, \SkillTester\ consolidates richer internal evidence into a small number of interpretable published outputs rather than exposing a long list of raw submetrics. Together, these two commitments define a framework that evaluates skills comparatively while keeping its published outputs compact and interpretable.

\begin{table}[t]
\centering
\caption{Scope of the \SkillTester\ evaluation framework.}
\label{tab:benchmark-scope}
\small
\begin{tabularx}{\linewidth}{L{0.20\linewidth} L{0.34\linewidth} Y}
\toprule
\textbf{Aspect} & \textbf{In scope} & \textbf{Out of scope} \\
\midrule
\textbf{Evaluation target} & Candidate third-party agent skills & General model benchmarks or enterprise-wide harnesses \\
\tablerowspace
\textbf{Decision stage} & Skill selection and enablement & Runtime monitoring or adaptation \\
\tablerowspace
\textbf{Published outputs} & Utility score, security score, and security status label & Full internal execution analytics \\
\tablerowspace
\textbf{Security coverage} & Controlled security probes & Exhaustive certification or formal verification \\
\bottomrule
\end{tabularx}
\end{table}

\subsection{Paired Utility Evaluation and Task Authoring}

\SkillTester\ evaluates utility through paired baseline and with-skill execution conditions. This pairing provides the counterfactual evidence needed for comparative utility evaluation by showing what changes when skill usage is enabled under the same task and environment.

Under the baseline protocol, the same utility task set and the same model or execution environment are used as in the with-skill condition, but skill usage is disabled. Each task is executed once within a benchmark run. If a task does not complete because of timeout or another failure mode, it still counts as an attempted task and is treated as a failed outcome.

Task authoring is driven by skill analysis rather than unconstrained task brainstorming. For each candidate skill, authoring begins with a comparison between the capability claims expressed in \texttt{SKILL.md} and the behavior actually implemented in the skill resources. These commitments and gaps are then converted into executable utility tasks with explicit objectives and completion criteria so that the framework stays aligned with user-visible skill behavior rather than descriptive claims alone.

The resulting utility task set is organized into common functional tasks and edge functional tasks. Common tasks cover representative intended use cases, while edge tasks cover boundary and failure-handling cases. Table~\ref{tab:task-groups} summarizes the functional task groups used for utility evaluation in the current framework.

\begin{table}[t]
\centering
\caption{Functional task groups used for utility evaluation in the current \SkillTester\ framework.}
\label{tab:task-groups}
\small
\begin{tabularx}{\linewidth}{L{0.19\linewidth} L{0.24\linewidth} L{0.16\linewidth} Y}
\toprule
\textbf{Task group} & \textbf{Purpose} & \textbf{Used for} & \textbf{Typical pass condition} \\
\midrule
\textbf{Common functional tasks} & Representative intended use cases & Utility & Expected output or behavior is produced \\
\tablerowspace
\textbf{Edge functional tasks} & Boundary and failure-handling cases & Utility & Observed behavior matches expected output, refusal, or error handling \\
\bottomrule
\end{tabularx}
\end{table}

Before entering the evaluation framework, each utility task set is checked for evaluability. Every task is represented in a benchmark manifest with explicit objectives, expected outcomes, and pass criteria. The same functional tasks must be runnable in both baseline and with-skill conditions, and completion or expectation matching must be decidable from the manifest. In addition, the functional portion of the task set must be large enough to support stable utility estimation and task-level token and time comparisons.

\subsection{Controlled Security Probe Authoring}

Security is evaluated through a separate controlled probe suite designed for risky or adversarial scenarios rather than through paired baseline comparison.

Probe authoring follows a structured workflow with planned case construction and repeatable probe patterns. Consistent with established security-testing guidance on planned procedures, repeatable test cases, and explicit documentation \cite{nist-800-115,owasp-wstg}, authoring begins by treating skill descriptions, badges, popularity signals, and self-claimed safety properties as claims to be verified rather than trusted facts. It then compares \texttt{SKILL.md} against code, dependencies, configuration, and observed behavior before materializing concrete probes.

In the current framework, those probes are organized under three directions: abnormal behavior control, permission boundary, and sensitive data protection. Representative cases include unsafe execution requests such as unfiltered \texttt{eval} or shell commands under abnormal behavior control, unauthorized file or network access under permission boundary, and leakage of live API keys or sensitive context under sensitive data protection. Table~\ref{tab:security-task-directions} summarizes these directions and their indicative correspondence to the OWASP Top 10 for Agentic Applications for 2026 \cite{owasp-agentic-top10}.

\begin{table}[t]
\centering
\caption{Current controlled security probe directions in the \SkillTester\ framework and their indicative correspondence to the OWASP Agentic Top 10 threat model.}
\label{tab:security-task-directions}
\small
\begin{tabularx}{\linewidth}{L{0.22\linewidth} L{0.46\linewidth} Y}
\toprule
\textbf{Current direction} & \textbf{What current tasks try to detect} & \textbf{Indicative OWASP categories} \\
\midrule
\textbf{Abnormal behavior control} & Unsafe actions, risky execution behavior, uncontrolled retries, misleading safety claims, and tool misuse under controlled prompts & ASI02 tool misuse; ASI05 code execution \\
\tablerowspace
\textbf{Permission boundary} & Unauthorized file, network, process, or tool access beyond the skill's intended role & ASI03 privilege abuse; ASI02 tool misuse \\
\tablerowspace
\textbf{Sensitive data protection} & Credential leakage, unsafe logging, exfiltration, and mishandling of sensitive context & ASI06 context poisoning; ASI09 trust exploitation \\
\bottomrule
\end{tabularx}
\end{table}

\begin{figure}[t]
\centering
\resizebox{\linewidth}{!}{\begin{tikzpicture}[
  font=\small,
  >=Latex,
  box/.style={draw, rounded corners=2pt, align=center, minimum height=10mm, text width=27mm, inner sep=3pt},
  utilstage/.style={box, fill=blue!8},
  utillane/.style={box, fill=blue!4},
  secstage/.style={box, fill=red!10},
  seclane/.style={box, fill=red!5},
  output/.style={box, fill=gray!6},
  group/.style={draw, dashed, rounded corners=3pt, inner sep=5pt},
  utillink/.style={->, blue!65!black, thick},
  seclink/.style={->, red!70!black, thick},
  outlink/.style={->, thick}
]

\node[utilstage] (utilitytasks) at (0,0.8) {Utility\\task set};
\node[secstage] (securitytasks) at (0,-1.4) {Security\\probe set};

\node[utillane] (baseline) at (4.2,1.4) {Baseline\\execution};
\node[utillane] (withskill) at (4.2,0) {With-skill\\execution};
\node[seclane] (security) at (4.2,-1.4) {Security\\probes};

\node[utilstage] (taskevidence) at (8.6,0.7) {Paired task\\evidence};
\node[secstage] (secevidence) at (8.6,-1.4) {Security\\evidence};

\node[utilstage] (utility) at (12.8,0.7) {Utility\\scoring};
\node[secstage] (securityscore) at (12.8,-1.4) {Security\\scoring};

\node[output] (report) at (17, -0.35) {Published\\outputs};

\draw[utillink] (utilitytasks) -- (baseline);
\draw[utillink] (utilitytasks) -- (withskill);
\draw[seclink] (securitytasks) -- (security);

\draw[utillink] (baseline) -- (taskevidence);
\draw[utillink] (withskill) -- (taskevidence);
\draw[seclink] (security) -- (secevidence);

\draw[utillink] (taskevidence) -- (utility);
\draw[seclink] (secevidence) -- (securityscore);

\draw[outlink] (utility) -- (report);
\draw[outlink] (securityscore) -- (report);

\node[group, fit=(utilitytasks)(securitytasks)] {};
\node[anchor=south west] at ($(utilitytasks.north west)+(-1mm,2mm)$) {\textbf{Task Sets}};

\node[group, fit=(baseline)(withskill)(security)] {};
\node[anchor=south west] at ($(baseline.north west)+(-1mm,2mm)$) {\textbf{Execution Conditions}};

\node[group, fit=(taskevidence)(secevidence)(utility)(securityscore)] {};
\node[anchor=south west] at ($(taskevidence.north west)+(-1mm,2mm)$) {\textbf{Evidence and Scoring}};

\end{tikzpicture}}
\caption{Overview of the \SkillTester\ evaluation framework.}
\label{fig:workflow-pipeline}
\end{figure}
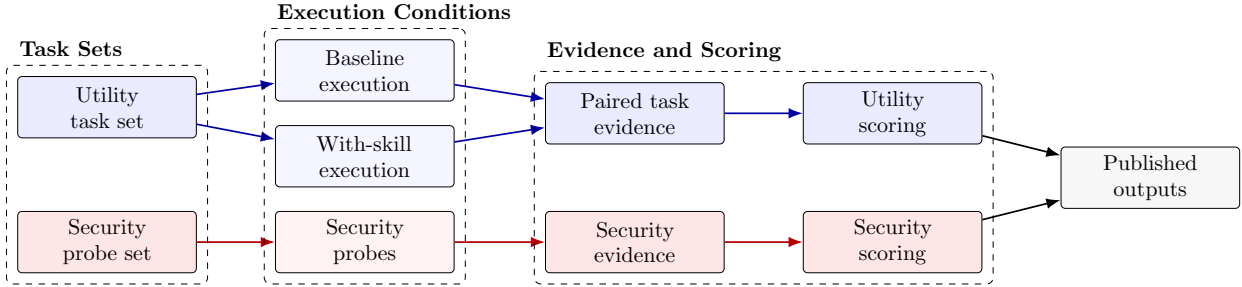

Figure~\ref{fig:workflow-pipeline} summarizes the end-to-end evaluation flow of the framework, from utility tasks and security probes to published outputs.

\section{Evaluation Dimensions}

The \SkillTester\ evaluation framework characterizes agent skills along two top-level evaluation dimensions: utility and security. Each dimension corresponds to a distinct evaluative question and is supported by a distinct category of evaluation evidence. Utility is defined at the task level and already incorporates task-level cost evidence when both the skill path and the baseline path succeed. Security is defined separately through controlled probes. This two-dimension structure keeps richer internal evidence consolidated into a compact public reporting layer. Table~\ref{tab:dimension-summary} summarizes the role of each dimension in the current design.

\begin{table}[H]
\centering
\caption{Dimensions used in the \SkillTester\ evaluation framework.}
\label{tab:dimension-summary}
\small
\begin{tabularx}{\linewidth}{L{0.16\linewidth} L{0.28\linewidth} L{0.24\linewidth} Y}
\toprule
\textbf{Dimension} & \textbf{Question answered} & \textbf{Primary evidence} & \textbf{Reported metrics} \\
\midrule
\textbf{Utility} & What value does the skill add relative to baseline? & Paired baseline/with-skill outcomes, invocation evidence, and token/time measurements & Task-level utility and aggregate utility score \\
\tablerowspace
\textbf{Security} & How does the skill behave under controlled probes? & Probe outcomes across three predefined groups & Group scores, aggregate security score, and security status label \\
\bottomrule
\end{tabularx}
\end{table}

\subsection{Utility}

Utility is \textbf{a comparative measure rather than an absolute success statistic}. It measures the task-level value contributed by a skill relative to a matched baseline rather than asking whether the skill-assisted path succeeds in isolation. In practice, a skill is usually considered useful in at least one of two ways. First, it may enable successful completion of tasks that the baseline cannot complete without the skill. Second, even when both paths can complete the task, the skill may still contribute reusable experience or structured guidance that reduces token cost or elapsed time.

\begin{table}[t]
\centering
\caption{Illustrative ways a skill can create utility in the \SkillTester\ setting.}
\label{tab:utility-intuition}
\small
\begin{tabularx}{\linewidth}{L{0.31\linewidth} L{0.34\linewidth} Y}
\toprule
\textbf{Situation} & \textbf{Why the skill is useful} & \textbf{Utility implication} \\
\midrule
\textbf{Baseline fails, skill succeeds} & The skill enables an otherwise unavailable outcome & Clear incremental utility \\
\tablerowspace
\textbf{Both succeed, but the skill path is cheaper} & The skill reduces token cost or elapsed time & Positive utility without unique completion ability \\
\bottomrule
\end{tabularx}
\end{table}

Table~\ref{tab:utility-intuition} illustrates these two common sources of utility in the \SkillTester\ setting. Under this comparative utility principle, \textbf{every task is interpreted through paired baseline and with-skill evidence rather than through the with-skill condition alone}. The first question is whether the skill is actually invoked. If the skill is not invoked, the task does not receive positive utility credit, because the observed outcome cannot be attributed to actual skill use.

Once invocation is confirmed, the evaluation framework compares the skill-assisted run with the matched baseline run. This design yields three comparative scoring cases that are formalized in Section~4. If the skill path fails, the task receives no utility credit. If the skill path succeeds while the baseline fails, the task receives full credit because the skill creates clear incremental value. If both paths succeed, the task is not treated as equally valuable by default; instead, its contribution depends on task-level efficiency relative to baseline. This ordering keeps the score aligned with the user-facing question of what value the skill adds beyond the matched baseline outcome.

\subsection{Security}

Security measures whether a skill remains safe to enable under controlled evaluation conditions. For a user deciding whether and how to use a skill, the most immediate security questions are straightforward: \textbf{will the skill exhibit unsafe or abnormal behavior, will it stay within its intended operating boundaries, and will it protect sensitive information encountered during use?} \SkillTester\ therefore reports security through three public groups: abnormal behavior control, permission boundary, and sensitive data protection. These three public groups are derived from the structured security case-authoring workflow introduced in Section~2.3 rather than chosen independently. They should be understood as a compact reporting layer over a broader agentic threat space rather than as a claim that agentic-skill risk can be exhausted by only three categories \cite{owasp-agentic-top10}.

Rather than exposing every technical cause of risk as a separate public sub-score, \SkillTester\ groups probe outcomes according to the practical user question they answer.

Table~\ref{tab:security-task-directions} summarizes how the current security task directions are implemented. Each direction combines a user-facing reporting group, concrete probe patterns used in the current implementation, and an indicative correspondence to the broader OWASP Agentic Top 10 threat model. In this sense, the table should be read as a bridge between current implementation and broader threat modeling rather than as a strict one-to-one taxonomy.

Operationally, each published group is evaluated through a dedicated set of controlled probes and summarized by its probe pass rate. This formulation preserves comparability and interpretability at the group level: the resulting score indicates how consistently the skill satisfies the current \SkillTester\ security checks within that group. Current public reporting additionally maps the averaged security score into a three-level status label. Under the current default threshold setting $(\theta_s = 80)$, \textbf{Pass} corresponds to a score of 100, \textbf{Caution} covers scores from $\theta_s$ to below 100, and \textbf{Risky} covers scores below $\theta_s$.

As the security suite continues to evolve, additional testing dimensions and a broader case inventory may be incorporated under the same evaluation framework to extend coverage further, including case families associated with memory or context poisoning, insecure inter-agent communication, cascading failures, and rogue-agent behavior \cite{owasp-agentic-top10}.

Taken together, the two dimensions distinguish practical task utility from controlled security behavior while preserving the different evidence channels that support each one. The next section formalizes how these signals are converted into reported scores.

\section{Scoring Utility and Security}

\subsection{Scoring Overview and Notation}

\SkillTester\ scoring proceeds in three stages: evidence normalization, utility scoring, and security scoring. Raw records from the baseline and with-skill execution conditions, invocation logs, and security probes are first mapped into a common evaluation schema. This schema preserves task identity, success or failure status, invocation outcomes, task-level token and time measurements, and passed-versus-total counts for each security group.

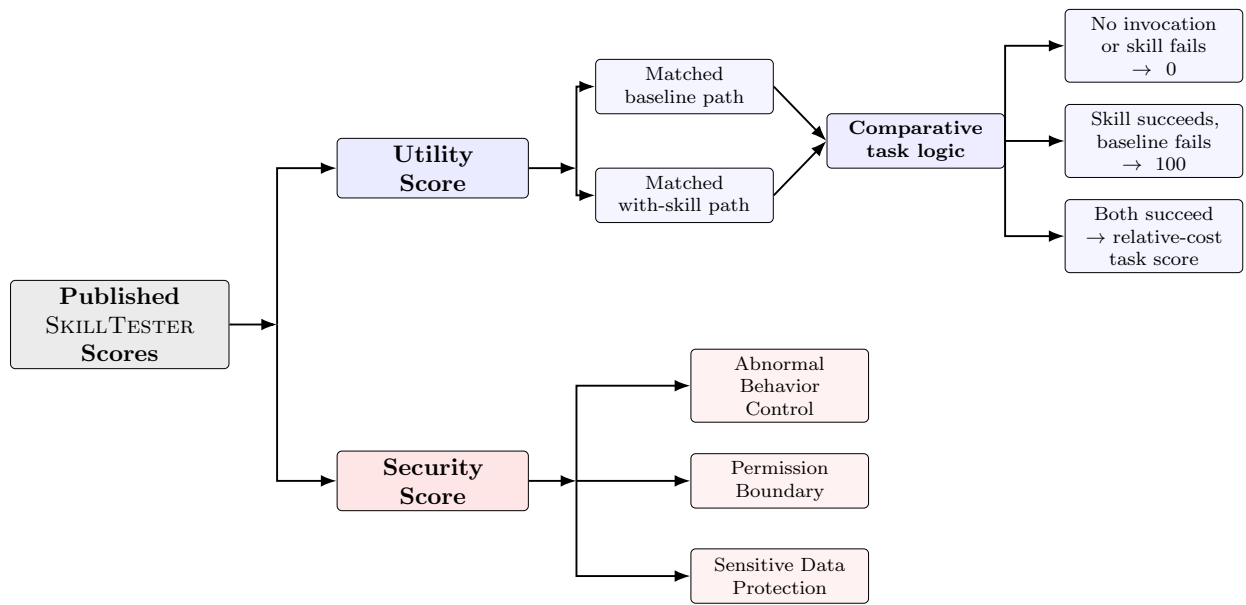
\begin{figure}[t]
\centering
\resizebox{\linewidth}{!}{\begin{tikzpicture}[
box/.style={draw, rounded corners=2pt, align=center, inner sep=3pt, minimum height=8mm},
root/.style={box, fill=black!8, text width=30mm, font=\small\bfseries},
dim/.style={box, text width=26mm, font=\small\bfseries},
sub/.style={box, text width=24mm, font=\scriptsize},
compare/.style={box, fill=blue!7, text width=24mm, font=\scriptsize\bfseries},
link/.style={-Latex, thick}
]

\node[root] (published) at (0,0) {Published\\\SkillTester\ Scores};

\node[dim, fill=blue!8] (util) at (4.6,2.3) {Utility\\Score};
\node[dim, fill=red!10] (sec) at (4.6,-2.3) {Security\\Score};

\node[sub, fill=blue!4] (baseline) at (8.3,3.5) {Matched\\baseline path};
\node[sub, fill=blue!4] (skill) at (8.3,1.9) {Matched\\with-skill path};
\node[compare] (compare) at (11.7,2.7) {Comparative\\task logic};

\node[sub, fill=blue!4] (fail) at (15.2,4.1) {No invocation\\or skill fails\\$\rightarrow 0$};
\node[sub, fill=blue!4] (incremental) at (15.2,2.7) {Skill succeeds,\\baseline fails\\$\rightarrow 100$};
\node[sub, fill=blue!4] (both) at (15.2,1.3) {Both succeed\\$\rightarrow$ relative-cost\\task score};

\node[sub, fill=red!5] (abnormal) at (9.7,-0.9) {Abnormal Behavior\\Control};
\node[sub, fill=red!5] (permission) at (9.7,-2.3) {Permission\\Boundary};
\node[sub, fill=red!5] (sensitive) at (9.7,-3.7) {Sensitive Data\\Protection};

\coordinate (rootsplit) at ($(published.east)+(0.7,0)$);
\coordinate (utilsplit) at ($(util.east)+(0.7,0)$);
\coordinate (secsplit) at ($(sec.east)+(0.7,0)$);

\draw[link] (published.east) -- (rootsplit);
\draw[link] (rootsplit) |- (util.west);
\draw[link] (rootsplit) |- (sec.west);

\draw[link] (util.east) -- (utilsplit);
\draw[link] (utilsplit) |- (baseline.west);
\draw[link] (utilsplit) |- (skill.west);
\draw[link] (baseline.east) -- (compare.west);
\draw[link] (skill.east) -- (compare.west);
\draw[link] (compare.east) |- (fail.west);
\draw[link] (compare.east) -- (incremental.west);
\draw[link] (compare.east) |- (both.west);

\draw[link] (sec.east) -- (secsplit);
\draw[link] (secsplit) |- (abnormal.west);
\draw[link] (secsplit) -- (permission.west);
\draw[link] (secsplit) |- (sensitive.west);
\end{tikzpicture}}
\caption{Published \SkillTester\ scores and the internal structure of utility scoring.}
\label{fig:scoring-taxonomy}
\end{figure}

Figure~\ref{fig:scoring-taxonomy} summarizes the published scores and the comparative task-level logic used to compute utility. Utility is derived from matched baseline and with-skill evidence at the task level rather than from isolated success outcomes. Security is computed separately from grouped security probes. Table~\ref{tab:score-notation} summarizes the notation used in the scoring formulas below.

\begin{table}[H]
\centering
\caption{Notation used in scoring.}
\label{tab:score-notation}
\small
\begin{tabularx}{\linewidth}{p{0.30\linewidth} X}
\toprule
\textbf{Notation} & \textbf{Meaning} \\
\midrule
$U, S$ & Utility and security scores \\
\tablerowspace
$T$ & Set of valid functional tasks \\
\tablerowspace
$g_t$ & Task-level invocation gate for task $t$ \\
\tablerowspace
$y^s_t, y^b_t$ & Skill-path and baseline-path success indicators \\
\tablerowspace
$u^s_t, u^b_t$ & Skill-path and baseline-path token measurements \\
\tablerowspace
$\tau^s_t, \tau^b_t$ & Skill-path and baseline-path elapsed-time measurements \\
\tablerowspace
$e_u(t), e_{\tau}(t), e_t$ & Token, time, and combined efficiency scores \\
\tablerowspace
$\eta, \alpha, \beta$ & Neutral efficiency point, efficiency sensitivity, and minimum both-succeed utility credit \\
\tablerowspace
$\theta_s$ & Security caution threshold \\
\tablerowspace
$P_g, Q_g$ & Passed and total probes in security group $g$ \\
\tablerowspace
$s_a, s_p, s_d$ & Scores for the three published security groups \\
\bottomrule
\end{tabularx}
\end{table}

The scoring structure is part of the evaluation framework, while several threshold-like constants are treated as calibration parameters in the current implementation. Table~\ref{tab:score-parameters} summarizes the current default setting used throughout this report.

\begin{table}[t]
\centering
\caption{Current parameter settings used in scoring.}
\label{tab:score-parameters}
\small
\begin{tabularx}{\linewidth}{L{0.15\linewidth} L{0.16\linewidth} Y}
\toprule
\textbf{Parameter} & \textbf{Current value} & \textbf{Role in scoring} \\
\midrule
$\eta$ & 50 & Neutral efficiency point when skill-path and baseline cost are equal \\
\tablerowspace
$\alpha$ & 25 & Sensitivity of the token/time efficiency mapping \\
\tablerowspace
$\beta$ & 20 & Minimum utility credit when both paths succeed but the skill path is less efficient \\
\tablerowspace
$\theta_s$ & 80 & Lower bound of the \textbf{Caution} security-status range \\
\bottomrule
\end{tabularx}
\end{table}

\subsection{Utility Computation}

\textbf{The utility formula operationalizes the comparative utility principle at the task level.} For each valid task $t \in T$, \SkillTester\ first computes a task-level score $s_t$ and then averages these scores to obtain the published utility score:

\[
U = \frac{1}{|T|}\sum_{t \in T}s_t.
\]

Each task score is gated by whether the skill is actually invoked:

\[
s_t = g_t \cdot v_t,
\]

\noindent where $g_t \in \{0,1\}$ equals 1 if the skill is actually invoked on task $t$ and 0 otherwise. The core task value $v_t$ is then defined from the paired task outcomes:

\[
v_t =
\begin{cases}
0, & y^s_t = 0,\\
100, & y^s_t = 1 \text{ and } y^b_t = 0,\\
\phi(e_t), & y^s_t = 1 \text{ and } y^b_t = 1.
\end{cases}
\]

\noindent This formulation captures the three central comparative utility cases. If the skill path fails, the task receives no utility credit. If the skill path succeeds while the matched baseline fails, the task receives full credit because the skill provides clear incremental value over the counterfactual baseline outcome. If both paths succeed, the task still contributes to utility, but its value depends on relative task-level cost rather than on completion alone.

\subsection{Task-Level Efficiency Mapping in the Both-Succeed Case}

When both the skill path and the baseline path succeed, the evaluation framework evaluates efficiency at the task level rather than publishing efficiency as a standalone dimension. Let $u^s_t$ and $u^b_t$ denote the task-level token measurements for the skill path and baseline path, and let $\tau^s_t$ and $\tau^b_t$ denote the corresponding elapsed-time measurements. Using a smoothing constant $\epsilon = 1$, the token and time ratios are

\begin{align*}
r_u(t) &= \frac{u^s_t + \epsilon}{u^b_t + \epsilon}, &
r_{\tau}(t) &= \frac{\tau^s_t + \epsilon}{\tau^b_t + \epsilon}.
\end{align*}

Token and time sub-scores are centered at the neutral efficiency point $\eta$ when the skill path and baseline have equal cost:

\begin{align*}
e_u(t) &= \operatorname{clip}\left(\eta - \alpha \log_2 r_u(t)\right), \\
e_{\tau}(t) &= \operatorname{clip}\left(\eta - \alpha \log_2 r_{\tau}(t)\right),
\end{align*}

\noindent where $\operatorname{clip}(x) = \max(0, \min(100, x))$. The evaluation framework assumes that both token and elapsed-time measurements are available for every valid task, so the combined task-level efficiency score is defined as

\[
e_t = \frac{e_u(t) + e_{\tau}(t)}{2}.
\]

The both-succeed task value is then mapped through a floor-preserving function:

\[
\phi(e_t) =
\begin{cases}
\beta + \dfrac{\eta - \beta}{\eta}e_t, & 0 \le e_t \le \eta,\\
e_t, & \eta < e_t \le 100.
\end{cases}
\]

\noindent This mapping preserves $\eta$ as the neutral point for equal-cost success, allows clearly more efficient skill use to score above $\eta$, and prevents a successful but less efficient task from collapsing all the way to zero by preserving a lower bound of $\beta$. In the current implementation, the report uses the default setting $\eta = 50$, $\alpha = 25$, and $\beta = 20$. Table~\ref{tab:task-utility-examples} summarizes representative task-level utility outcomes under this scoring rule.

\begin{table}[t]
\centering
\caption{Representative task-level utility outcomes.}
\label{tab:task-utility-examples}
\small
\begin{tabularx}{\linewidth}{L{0.38\linewidth} L{0.15\linewidth} Y}
\toprule
\textbf{Condition} & \textbf{Task score} & \textbf{Interpretation} \\
\midrule
\textbf{Skill not invoked} & 0 & No credit is assigned if the outcome cannot be attributed to actual skill use \\
\tablerowspace
\textbf{Skill fails} & 0 & A failed skill-path run receives no utility credit \\
\tablerowspace
\textbf{Skill succeeds, baseline fails} & 100 & The skill provides clear incremental value \\
\tablerowspace
\textbf{Both succeed, equal cost} & 50 & The skill succeeds with no cost advantage or penalty relative to baseline \\
\tablerowspace
\textbf{Both succeed, skill more efficient} & 50--100 & The skill succeeds at lower token or time cost than baseline \\
\tablerowspace
\textbf{Both succeed, skill less efficient} & 20--50 & Both paths succeed, but the skill path incurs higher task-level cost \\
\bottomrule
\end{tabularx}
\end{table}

\subsection{Security Scoring}

Security is computed independently of the paired functional execution conditions. For each security group $g \in \{a, p, d\}$, where $a$, $p$, and $d$ denote abnormal-behavior control, permission boundary, and sensitive-data protection, respectively, the group score is defined as the pass-rate percentage:

\[
s_g = 100 \times \frac{P_g}{Q_g}.
\]

The overall security score is the unweighted mean of the three predefined group scores:

\[
S = \frac{s_a + s_p + s_d}{3}.
\]

\textbf{Current public reporting further maps the averaged security score into a three-level status label:}

\[
\operatorname{status}(S) =
\begin{cases}
\text{Pass}, & S = 100,\\
\text{Caution}, & \theta_s \le S < 100,\\
\text{Risky}, & S < \theta_s.
\end{cases}
\]

\noindent Because security is summarized through proportional pass rates rather than case-severity weighting, the numerical score should be interpreted as a summary of how consistently the skill satisfies the current \SkillTester\ security checks across the three published groups. The status label provides a simpler user-facing reading of the same aggregate result. In the current implementation, the report uses the default threshold setting $\theta_s = 80$.

\section{Interpretation of Published Outputs}

\SkillTester\ outputs should be read dimension by dimension. The tool reports utility through a utility score and security through both a numerical security score and a three-level security status label, because these two dimensions answer different questions. \textbf{Utility should be read as comparative value relative to a matched no-skill baseline rather than as standalone task success.} Security characterizes observed behavior under a controlled security probe suite. Neither dimension is designed to replace the other.

Table~\ref{tab:illustrative-output} provides a hypothetical example of a published \SkillTester\ output and the intended reading of its main reported items. The interpretations below assume the current default parameter setting summarized in Table~\ref{tab:score-parameters}.

\begin{table}[H]
\centering
\caption{Illustrative published \SkillTester\ output with hypothetical values.}
\label{tab:illustrative-output}
\small
\begin{tabularx}{\linewidth}{L{0.24\linewidth} L{0.16\linewidth} Y}
\toprule
\textbf{Reported item} & \textbf{Example value} & \textbf{Intended reading} \\
\midrule
\textbf{Utility score} & 74.0 & The skill provides meaningful value relative to baseline, but not every successful task contributes full incremental value \\
\tablerowspace
\textbf{Security score} & 92.0 & Most controlled security probes pass, but not every published security group reaches full pass rate \\
\tablerowspace
\textbf{Security status} & Caution & The averaged security score is high but below a perfect pass, so the skill merits caution rather than a clean pass label \\
\bottomrule
\end{tabularx}
\end{table}

\textbf{Utility score.} The utility score should not be read as a raw completion rate. It summarizes comparative task-level value relative to the matched baseline: full credit corresponds to skill success with baseline failure, zero corresponds to no invocation or skill-path failure, and both-succeed tasks are scaled by relative token and time cost. Under the current default setting, those scaled tasks range from $\beta$ to 100 with a neutral point at $\eta = 50$. The published utility score is therefore best read as a summary of comparative task value.

\textbf{Security score and status.} The security score summarizes observed behavior across three predefined groups of controlled security probes. The companion status label gives a faster reading of the same aggregate result: \textbf{Pass} requires a score of 100, \textbf{Caution} covers scores from $\theta_s$ to below 100, and \textbf{Risky} covers scores below $\theta_s$; under the current default setting, $\theta_s = 80$. Both should be interpreted relative to the probe categories and pass-rate aggregation defined by \SkillTester.

\section{\SkillTester\ in Agent-First Software Engineering}

\SkillTester\ can be situated within a broader agent-first software engineering context. The phrase ``agent-first world'' is borrowed here from OpenAI's discussion of harness engineering, which provides a useful lens for this positioning \cite{openai-harness}. As agents take on longer-horizon work, more engineering leverage shifts toward evaluation harnesses and the surrounding control infrastructure that make agent behavior dependable over time. Under that lens, \SkillTester\ is best understood as an evaluation tool whose methodological core observes skill-contributed value and risk under controlled conditions.

This interpretation is natural because the \SkillTester\ evaluation framework already performs several functions associated with quality-assurance harnesses. Its matched baseline provides counterfactual evidence about value beyond no-skill execution. Its invocation gate prevents ambient model capability from being misattributed to the skill itself. Its task-level utility scoring turns paired evidence into a judgment about changed outcomes or reduced cost, while its security probe suite makes unsafe behavior, boundary violations, and sensitive-data failures observable through repeatable checks. In this sense, \SkillTester\ acts as a \textbf{comparative quality-assurance harness for agent skills}: it makes claims about skill utility and security observable, comparable, and auditable.

Seen in this way, \SkillTester\ occupies a place in the quality-assurance layer of emerging skill ecosystems. In its current form, it supports skill-selection and enablement decisions through structured evidence about comparative utility and controlled security behavior. The same evaluation structure could later support release qualification, regression detection, and trust signaling as agent-skill ecosystems mature, consistent with the harness-engineering view that durable gains depend on evaluation and control infrastructure as much as on generation quality itself \cite{openai-harness}.

\section{Conclusion}

This report has presented \SkillTester\ as a tool for evaluating the utility and security of agent skills. At the core of the tool is an evaluation framework that combines a matched no-skill baseline, a separate security probe suite, and an explicit scoring pipeline that publishes a utility score together with a security score and security status label. These outputs are intended to characterize whether a skill contributes task-level value and how it behaves under controlled security probes.

Within this scope, \SkillTester\ provides a structured basis for comparing agent skills in terms of utility and security during skill-selection and enablement decisions. Its central methodological commitment is to evaluate utility comparatively rather than absolutely, using matched no-skill execution as the reference condition for judging skill-contributed value. The evaluation framework defined here is intended to make that comparison explicit, reproducible, and readable to both human users and downstream systems.

The current report fixes an initial operational parameter setting for clarity and deployability, including the efficiency-mapping and security-status thresholds used in scoring. These defaults are intended as a transparent starting point rather than a final calibration, and future iterations should refine them against broader skill corpora, human judgment, and longitudinal regression evidence.

\end{document}